\documentclass[letterpaper]{JINST}
\title
{Solvable Examples of Drift and Diffusion of Ions in Non-uniform Electric Fields}
\author{
R. N. Cahn \\
Lawrence Berkeley National Laboratory\\
Berkeley, CA, 94720, USA\\E-mail: \email{rncahn@lbl.gov}}
\author{
J. D. Jackson\\
Lawrence Berkeley National Laboratory\\
Berkeley, CA, 94720, USA\\
E-mail: \email{jdjackson@lbl.gov}}
\abstract{
  The drift and diffusion of a cloud of ions in a fluid are distorted
  by an inhomogeneous electric field.  If the electric field carries
  the center of the distribution in a straight line and the field
  configuration is suitably symmetric, the distortion can be
  calculated analytically.  We examine the specific examples of fields
  with cylindrical and spherical symmetry in detail assuming the ion
  distributions to be of a generally Gaussian form.  The effects of
  differing diffusion coefficients in the transverse and longitudinal
  directions are included.
}


\usepackage{url}
\def\begeq{\begin{equation}}
\def\endeq{\end{equation}}
\def\begeqar{\begin{eqnarray}}
\def\endeqar{\end{eqnarray}}
\def\begit{\begin{itemize}}
\def\endit{\end{itemize}}
\def\non{\nonumber\\}

\def\bfr{{\bf r}}

\def\bfv{{\bf v}}

\def\bfE{{\bf E}}

\def\xhat{{\bf \hat x}}
\def\rminusrzero{{\rho}}

\begin{document}
 \section{Introduction}

 We address the time and spatial evolution of the ionization produced
 along a charged particle's track in a fluid (gas or liquid) subjected
 to a non-uniform, time-independent electric field. The prototypical
 calculation is for a localized ion density produced at some point
 ${\bf R}_{0}$ at time $t = 0$, described by a delta function $
 \delta({\bf r} - {\bf R}_{0})$. Of interest is the diffusion in both
 the direction of the electric field and transverse to it, which
 governs the distributions in arrival time and transverse position at
 some detecting surface. We focus on a volume that is bounded by
 cylindrical electrodes that produce a radial electric field, but
 begin with more general considerations.  Spherical geometry is
 considered in an appendix.

 The motion of electrons and atomic or molecular ions in
 electromagnetic fields has a huge experimental and theoretical
 literature. See, for example, the texts and cited literature of
 Huxley and Crompton\cite{Huxley} and Mason and McDaniel\cite{Mason}.
 An extensive literature about diffusion of charged particles in
 plasmas also exists, but conditions there are far from our concern of
 relatively few ions moving in a neutral medium.


 We make the simplifying assumptions that the ion mobility $\mu $ and
 the longitudinal and transverse diffusion constants, $D_{L}$ and
 $D_{T}$, are independent of electric field strength $ E $ or rather
 of $ E/N $, where $ N $ is the particle number density of the
 fluid. The data compilations of transport properties of gaseous ions
 \cite{Ellis_1, Ellis_2, Ellis_3} show that the mobilities of ions in
 noble gases are field-independent (to $ \pm 10 \ \% $) typically for
 $E/N < 60-100 \ {\rm Td}$.\cite{notation} The longitudinal
 diffusion constants $ D_{L} $ for ions in noble gases are similarly
 roughly field-independent typically for $ E/N < 10-2 0 \ {\rm Td}
 $. For this range of $ E/N $, the ratio of the diffusion constants
 is $ D_{L}/D_{T} \approx 1$; only at higher values of $ E/N $ is $
 D_{L}/D_{T} $ greater than unity (and both are field-dependent). See, for example,
 figure 5-2-4 of Ref.~\cite{Mason}.
  
 Within our assumptions, we hypothesize a regime in which the
microscopic (atomic) processes that establish the local drift
velocity and diffusion parameters occur on length and time scales
small compared to the changes in magnitude or direction of the
electric field or the average drift position.  Such a regime permits
the use of the simple diffusion equation rather than the full
Boltzmann equation.

 For electrons in gases, field-independent mobilities occur typically
only for $ E/ N < 10^{-2} \ {\rm Td}$,with various field
dependencies at larger $ E/N $.\cite{Huxley} Diffusion constants for
electrons often exhibit rapidly varying field-strength dependence
because of resonant behaviour in the momentum transfer cross sections. Our approximations seem to have limited applicability for
electrons.

\section{Preliminaries and initial Ansatz}

Ions in a fluid and subject to an electric field can be described by a
charge density $\psi(\bfr,t)$ satisfying the diffusion equation

\begeq
\frac {d\psi}{dt}=\frac{\partial\psi}{\partial t}+{\bf
  v}\cdot{\bf\nabla}\psi=D\nabla^2\psi.\label{eq:one}
\endeq

We take the velocity ${\bf v}$ to be given by the mobility $\mu$ times
the electric field ${\bf E}$.  We imagine that from the prescribed
electric field and mobility we can calculate the classical trajectory
$\bfr_0(t)$ of a point charge located at $\bfr={\bf R}_0$ at time
$t=0$.  We indicate the velocity along the path by ${\bf v}_0(t)$ and,
in particular,
\begeq
\bfv_0(t)=\mu \bfE ({\bfr}_0(t)).
\endeq
The trajectory is given implicitly as
\begeq
{\bf r}_0(t)={\bf R}_0 +\int_0^tdt' \mu {\bf E}(r_0(t')).\label{eq:three}
\endeq
We are seeking a solution proportional to $\delta({\bfr} -{\bf R}_0)$
at $t=0$.  We begin with an Ansatz for $\psi$:

\begeq
\psi=\frac {[\det(I+g)]^{1/2}}{(4\pi Dt)^{3/2}}\exp\left\{-[\bfr-\bfr_0(t)]_i[\delta_{ij}+g(t)_{ij}][\bfr-\bfr_0(t)]_j/(4Dt)\right\}.\label{eq:ansatz}
\endeq
The factor $[\det(I+g)]^{1/2}$ ensures that $\psi$ is normalized properly:

\begeq
\int d^3\bfr \, \psi =1,
\endeq
which must be the case if we solve the initial equation, which derives from the continuity equation.  In the absence of the electric field, the familiar solution is given by Eq.(\ref{eq:ansatz}) with $g_{ij}=0$ and $\bfr_0(t)={\bf R}_0$.
The basic assumption is that the form is Gaussian with respect to the difference between the point at which the density is evaluated and the location of the classical trajectory at that time, $\bfr_0(t)$.  The form is motivated by the observation that, not only in the absence of the electric field, but also for a constant electric field, the solution is given by Eq.(\ref{eq:ansatz}) with $g_{ij}=0$ . 
 We calculate the time and spatial derivatives of $\psi$,  writing $N(t)$ for $[\det(I+g)]^{1/2}$ and 
 $\rho_i$ for $(\bfr-\bfr_0(t))_i$:
\begeqar
4Dt\frac{\partial\psi}{\partial t}&=&{\rminusrzero_i(\delta_{ij}+g_{ij})\rminusrzero_j}\frac\psi t -6D\psi +{2\rminusrzero_i (\delta_{ij}+g_{ij})v_{0j}}\psi \non
&&\qquad-{\rminusrzero_i\frac{dg_{ij}}{dt}\rminusrzero_j}\psi
+4Dt\frac {d\ln N}{dt}\psi;\\
4Dt\nabla_k\psi&=&-{2(\delta_{jk}+g_{jk})\rminusrzero_j}\psi;\\
4D^2t\nabla_k\nabla_k \psi&=&(\delta_{kj}+g_{kj})\rminusrzero_j(\delta_{kl}+g_{kl})\rminusrzero_l\frac\psi t-2(\delta_{kk}+g_{kk})D\psi\non
&=&{\rminusrzero_j(\delta_{jl}+2g_{jl}+g_{kj}g_{kl})\rminusrzero_l}\frac\psi t
-2(3+g_{kk})D\psi;\\
4Dt\mu\bfE\cdot{\bf \nabla}\psi&=&-{2\mu E_k(\delta_{kj}+g_{jk})\rminusrzero_j}\psi.
\endeqar
We now expand $\bfE(\bfr)$ around $\bfr_0(t)$, keeping first order terms,
\begeq
\bfv_0(t)-\mu \bfE(\bfr)=\mu(\bfE(\bfr_0(t))-\bfE(r))=-\mu\rminusrzero_i\partial_i\bfE(\bfr_0(t))+\ldots.
\endeq
The omitted terms in the expansion are of order $ (\rho/L)^{n}, \ n \geq 1$, times the first order term, where $L$ is the length scale over which the electric field changes appreciably. If the spread of the diffusion, $ O(\rho_{max})$, is small compared to $ L $, our approximation should be reliable.
We combine these into Eq.(\ref{eq:one})  to find the equations for $g_{ij}$:
\begeqar
&&\rminusrzero_i\frac{(\delta_{ij}+g_{ij})}t\rminusrzero_j
 -{6D}+4Dt\frac {d\ln N}{dt}\psi
-\rminusrzero_i\frac{dg_{ij}}{dt}\rminusrzero_j\non
&&\qquad\qquad=\rminusrzero_j\frac{(\delta_{jl}+2g_{jl}+g_{kj}g_{kl})}t\rminusrzero_l
-2(3+g_{kk})D\non
&&\qquad\qquad\qquad+2\mu\rminusrzero_i\partial_i E_k(\delta_{kj}+g_{jk})\rminusrzero_j.
\endeqar
The terms with and without quadratic dependence on $\rminusrzero$ must cancel separately.  In fact, we find that the equations that derive from terms with quadratic dependence ensure that the terms without that dependence cancel, as well.
The terms without the quadratic dependence give
\begeq
1-\frac{2t}{3}\frac{d\ln N}{dt}=1+\frac 13g_{kk}\label{eq:no-quad}
\endeq
The terms with quadratic dependence on $\rminusrzero$ give
\begeqar
&&\frac{(\delta_{ij}+g_{ij})}t
-\frac{dg_{ij}}{dt}=\frac{(\delta_{ji}+2g_{ji}+g_{kj}g_{ki})}t
+2\mu\partial_i E_k(\delta_{kj}+g_{jk})
\endeqar
 or
 \begeq
 \frac{dg_{ij}}{dt}+\frac{g_{ij}+g_{ik}g_{kj}}t + 2\mu\partial_iE_k(\delta_{kj}+g_{kj})=0\label{eq:b}
 \endeq
 \section{Restriction to rectilinear motion and symmetric geometry}
 Only the symmetric part of $g_{ij}$ enters $\psi$.  Consistency requires that the antisymmetric part of Eq.(\ref{eq:b}) vanish:
 
 \begeq
 (\delta_{ij}+g_{ij})\partial_kE_j=(\delta_{kj}+g_{kj})\partial_i E_j
 \endeq
 The vanishing of the curl of ${\bf E}$ is enough to take care of the $\delta_{ij}$ portion but we are left with the requirement
 \begeq
 g_{ij}\partial_k E_j=g_{kj}\partial_i E_j\label{requirement}
 \endeq
 
 In general, the product of two symmetric matrices is not symmetric.  We need to impose additional constraints.  In particular, we suppose that the motion is rectilinear along the $x$ direction, i.e.
 $\bfr(t)=r(t)\xhat$, ${\bf E}(t)=E_x(t)\xhat$. Write the electrostatic potential in the neighborhood of the line of motion as an expansion in $y$ and $z$:
 \begeqar
 \Phi(x,y,z)&=& \Phi(x,0,0) + y\Phi_y(x) +z\Phi_z(x) \non
&&+\frac 12 y^2 \Phi_{yy}(x) +\frac 12z^2\Phi_{zz}(x)\non
&& +xy\Phi_{xy}(x) + xz\Phi_{xz}(x)+yz\Phi_{yz}(x)+\ldots\non\label{eq:expansion}
 \endeqar
 
 By hypothesis the electric field along $y=0,\ z=0$ is entirely in the $x$ direction. Then
 $\Phi_y(x)=\Phi_z(x)=\Phi_{xy}(x)=\Phi_{xz}(x)=0$ and Eq.(\ref{eq:expansion}) becomes, to second order inclusive,
 \begeqar
 \Phi(x,y,z)&=& \Phi(x,0,0) 
+\frac 12 y^2 \Phi_{yy}(x) +\frac 12z^2\Phi_{zz}(x)
+yz\Phi_{yz}(x)\non
 \endeqar
 
\begin{figure}\begin{center}
 \includegraphics[width=4in]{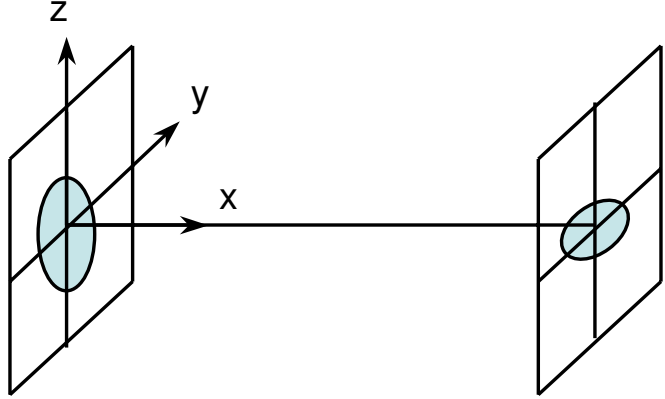}
 
 \caption[99]{The Ansatz for $\psi$, Eq.(\ref{eq:ansatz}), is applicable
   for rectilinear trajectories if the matrix $\partial_i E_j$ is
   diagonal, for a suitable choice of coordinates.  Taking $x$ along
   the direction of motion, a charge distribution (or potential)
   symmetric under $y \to -y$ and $z\to -z$ will ensure this.  The
   figure shows such a trajectory and charge distribution.  Here only
   the trajectory shown has the required symmetry.  For the
   cylindrical and spherical geometries all radial trajectories have
   the symmetry required by our Ansatz.}\label{fig:drift1}
\end{center}
\end{figure}

 If the surface charge distributions determining the electric field
 are symmetric under $y\to -y$ and $z\to -z$, the term $yz\Phi_{yz}$
 vanishes along with $\Phi_y,\Phi_z,\Phi_{xy}$ and $\Phi_{xz}$, and the
 only non-zero values of $\partial_iE_j$ along the line $(x,0,0)$
 occur for $i=j$.  Two clear examples are cylindrical symmetry with
 the $z$ direction taken as the axis of the cylinder and spherical
 symmetry.  In the former case all derivatives with respect to $z$
 vanish so $\partial_z E_z=0$ and in the latter, the derivatives with
 respect to $y$ and $z$ are equal so $\partial_zE_z=\partial_y E_y$.
 See figure \ref{fig:drift1}.
 
   Under any of these circumstances we can write 
 \begeq
 \partial_i E_j=C_i(x)\delta_{ij}
 \endeq
 and we can satisfy Eq.(\ref{requirement}) with the ansatz
 \begeq
 g_{ij}=g_i\delta_{ij}
 \endeq
 i.e. both $g_{ij}$ and $\partial_i E_j$ are diagonal along the linear trajectory path of Eq.(\ref{eq:three}).  Now Eq.(\ref{eq:b}) simplifies to
 \begeq
  \frac{dg_{i}}{dt}+\frac{g_{i}(1+g_{i})}t + 2\mu\partial_iE_i(1+g_{i})=0\label{eq:c}
 \endeq  
 where there is no summation on $i$.
\section {Example of cylindrical geometry}

For a cylindrical geometry, uniform in the $z$ direction and having azimuthal symmetry, with a radial electric field, we have 
\begeq
\bfE = \frac Ar {\bf\hat r},\label{eq:field}
\endeq
where $\bfr = x{\bf\hat x}+y{\bf\hat y}$ is now two-dimensional.  We assume here that the signs of the charge and field are such that the force is radially outward.  

We can anticipate the effect of the inhomogeneous electric field,
Eq.(\ref{eq:field}).  The portion of the cloud that is further out
than the center of the cloud experiences a weaker field and thus moves
more slowly than the outwardly moving center, while the portion of the
cloud closer in to the central axis experiences a greater field and
moves more rapidly than moving center.  Thus the spread of the cloud
in the radial direction should be smaller than in the absence of the
electric field. On the other hand, the diverging electric field will
tend to spread the cloud transversely, leading to a larger dispersion
than in the absence of the electric field.

Take the initial position ${\bf R}_0=(x_0,0,0)$.  The motion is along the $x$ axis where

\begeq
\partial_xE_x=-\frac A{x^2};\quad \partial_y E_y=\frac A{x^2};\quad \partial_x E_y=0=
\partial_yE_x,
\endeq
satisfying the symmetry requirements for our Ansatz.  Thus we have the diagonal form $\partial_i E_j=C_i(x)\delta_{ij}$ 
where
\begeqar
C_1(x)&=&-\frac A{x^2},\\
C_2(x)&=& \frac A{x^2},\\
C_3(x)&=&0.
\endeqar

 The trajectory is given by $\bfr_0(t)=r_0(t){\hat{\bf x}}$, with
\begeq
r_0(t)^2=x_0^2+2\mu A t,
\endeq
so that
\begeq
2\mu \partial_xE_x=-\frac{2\mu A}{x_0^2+2\mu A t}=-2\mu \partial_yE_y.
\endeq
  Introduce
\begeq
T=\frac{x_0^2}{2\mu A}
\endeq
so that
\begeq
r_0(t)=x_0\sqrt{1+\frac tT}.\label{eq:r_of_t}
\endeq
In the characteristic time $T$ the diffusion center would go a
distance $x_0/2$ if the field had the constant value $E=A/x_0$.  When
$t=3T$, the classical particle is twice as far from the axis  of the
cylinder in the $1/r$ field as it was at $t=0$.  From Eq.(\ref{eq:c}):

\begeqar
\frac{dg_1}{dt}&=&-\frac{g_1(1+g_1)}t+\frac 1{T+t}(1+g_1);\non
\frac{dg_2}{dt}&=&-\frac{g_2(1+g_2)}t-\frac 1{T+t}(1+g_2);\non
\frac{dg_3}{dt}&=&-\frac{g_3(1+g_3)}t.\non
\endeqar
Finally, writing $t=Ts$ 
\begeqar
\frac{dg_1}{ds}&=&-\frac{g_1(1+g_1)}s+\frac 1{1+s}(1+g_1);\non
\frac{dg_2}{ds}&=&-\frac{g_2(1+g_2)}s-\frac 1{1+s}(1+g_2);\non
\frac{dg_3}{ds}&=&-\frac{g_3(1+g_3)}s.\non\label{eq:nonlinear}
\endeqar
The equation for $g_3$ has the solution 

\begeq
g_3=\frac 1{Bs-1}
\endeq
where $B$ is a constant.  When $s=0$, all the $g_i$ are zero.  It follows that $B=\infty$ and $g_3$ remains zero: Diffusion in
the $z$ direction is unchanged by the presence of the radial electric field.

\begin{figure}\begin{center}
\includegraphics[width=4in,angle=90]{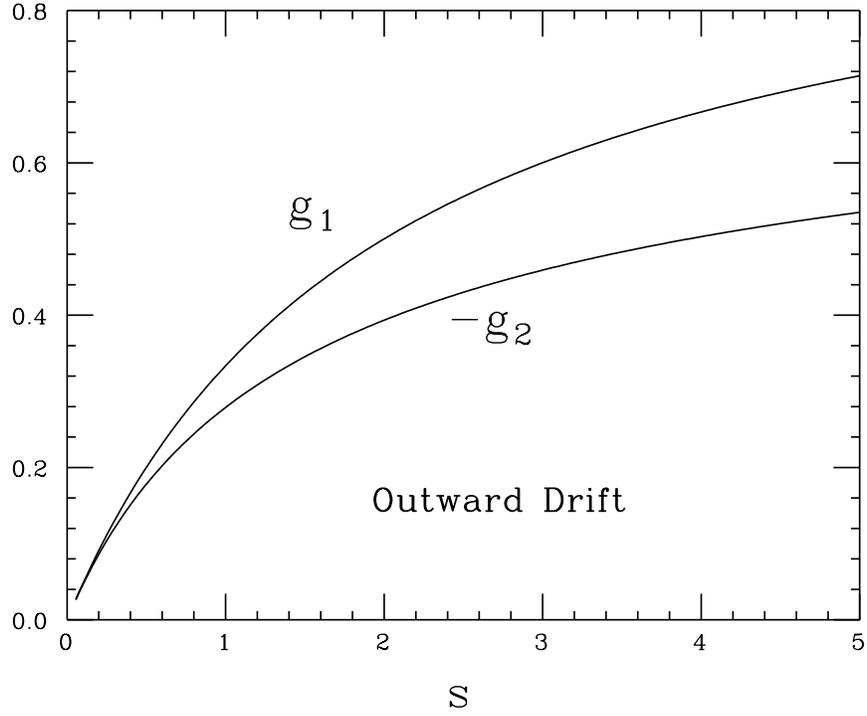}
\caption{The functions appearing in the Gaussian description of the charge density $\psi$ for outward drift, as a function of the dimensionless time variable $s=t/T$ for cylindrical geometry.}\label{fig:g1g2}\end{center}
\end{figure}

Eqs.(\ref{eq:nonlinear}) for $g_1$ and $g_2$ are Ricatti equations and  can be solved analytically.
The details of the solutions are given in Appendix A.  The results are
\begeqar
g_1&=& \frac s{2+s}\\
g_2&=&-1 +\frac s{(1+s)\ln(1+s)}
\endeqar
The behavior of the functions $g_1$ and $g_2$ is shown in figure \ref{fig:g1g2}.

\begin{figure}\begin{center}
\includegraphics[width=4in,angle=90]{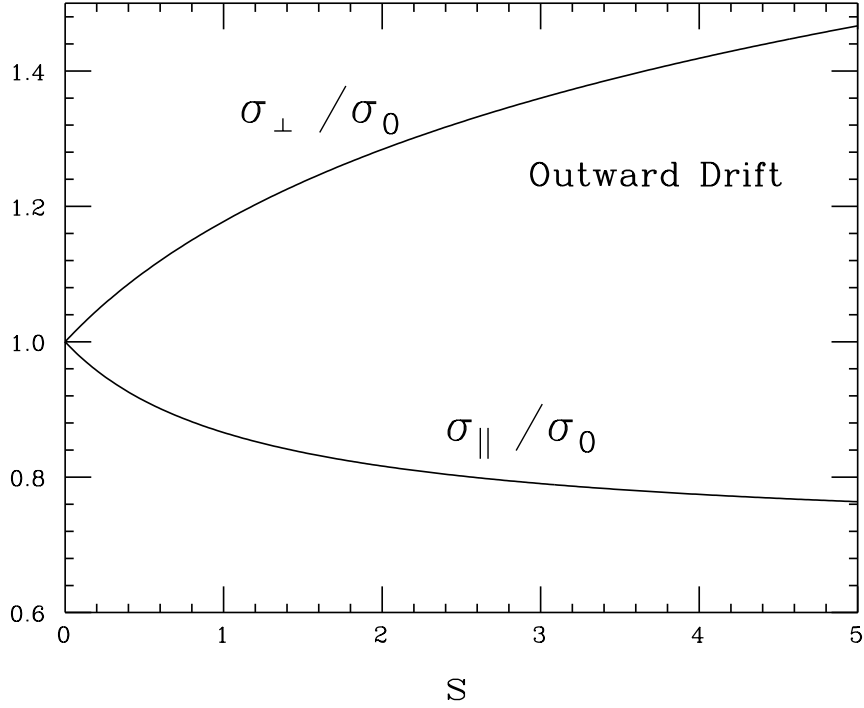}
\caption[99]{The effective dimensionless widths Eq.(\ref{eq:widths}) for outward drift, in the directions parallel and perpendicular to the radial direction as functions of the dimensionless time variable $s=t/T$ for cylindrical geometry.  If the diffusion constants differ in the longitudinal (parallel to the field) and transverse directions, multiply the ratio $\sigma_\parallel/\sigma_0$ by $\sqrt{D_L/D_T}$ and take $\sigma_0=\sqrt{2D_T t}$. (see Sec. 6).}\label{fig:sigmas}\end{center}
\end{figure}

The effective parallel and transverse spreads relative to $\sigma_0=\sqrt{2Dt}$ are
\begeq
\frac{\sigma_\parallel}{\sigma_0}=\sqrt{\frac {1}{1+g_1}};\quad 
\frac{\sigma_\perp}{\sigma_0}=\sqrt{\frac{1}{1+g_2}},\label{eq:widths}
\endeq
shown in figure \ref{fig:sigmas}.  In the $z$-direction the spread is $\sigma_3=\sigma_0$.
\clearpage
\section {Inward drift}
We have imagined that the radial drift is outward, towards larger $r$.  If the sign of the electric field is reversed, the derivation goes through much the same way except that
the sign of $T$ is negative.  Since $t=Ts$, positive times then correspond to negative values of $s$.  The center point of the diffusion cloud arrives at the central axis at $s=-1$.  Because the field increases in the direction of motion, we expect the effect of the field to be opposite to what we found above:  the dispersion should be greater in the radial direction and smaller in the transverse direction.  

\begin{figure}\begin{center}
\includegraphics[width=4in,angle=90]{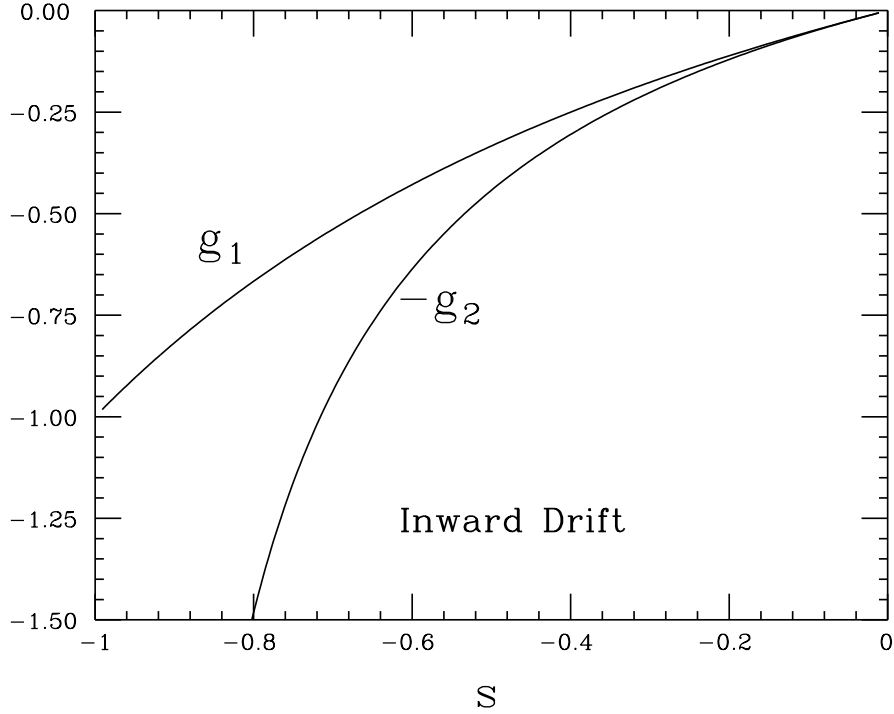}
\caption{The functions appearing in the Gaussian description of $\psi$ for inward drift in the cylinder. Negative values of $s$ correspond to positive values of $t$ according to $t=Ts$.  For $s=-1$, $\bfr_0(t)=0$.}
\label{fig:inward}\end{center}
\end{figure}

\begin{figure}\begin{center}
\includegraphics[width=4in,angle=90]{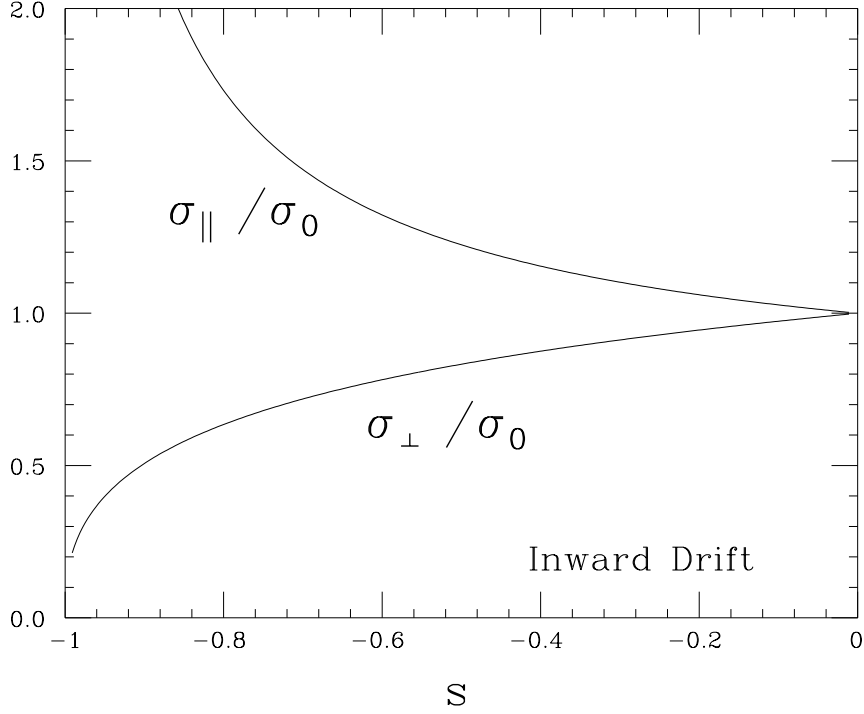}
\caption[99]{The effective dimensionless widths Eq.(\ref{eq:widths}) in
  the directions parallel and perpendicular to the radial direction
  for inward drift in the cylinder. Negative values of $s$ correspond
  to positive values of $t$ according to $t=Ts$.  If the diffusion
  constants differ in the longitudinal (parallel to the field) and
  transverse directions, multiply the ratio $\sigma_\parallel/\sigma_0$ by $\sqrt{D_L/D_T}$ and take $\sigma_0=\sqrt{2D_T t}$ (see
  Sec. 6).}\label{fig:inward-sigmas}\end{center}
\end{figure}

We show the results obtained for inward drift in Figs. \ref{fig:inward} and \ref{fig:inward-sigmas}.

\section{Different coefficients of diffusion for longitudinal and transverse directions}

It is straightforward to include the effect of having different diffusion coefficients $D_L$ and $D_T$ for the longitudinal and transverse directions.  For the Ansatz we take

\begeq
\psi=\frac {[\Pi_j(1+g_j)]^{1/2}}{(4\pi t)^{3/2}D_TD_L^{1/2}}\exp\left\{-\sum_i\frac{\rho_i^2(1+g_i)}{4D_it}\right\}.\label{eq:ansatztwo}
\endeq
with $D_1=D_L, D_2=D_3=D_T$.  Exactly the same differential equations emerge, but setting $\sigma_0=\sqrt{2D_Tt}$ we have
\begeqar
\frac{\sigma_\parallel}{\sigma_0}&=&\sqrt{\frac{D_L}{D_T}}\frac 1{\sqrt{1+g_1}}\\
\frac{\sigma_\perp}{\sigma_0}&=&\frac 1{\sqrt{1+g_2}}\\
\frac{\sigma_3}{\sigma_0}&=&1
\endeqar
\section{Diffusion of a track segment}

\begin{figure}\begin{center}
\includegraphics[width=2.5in]{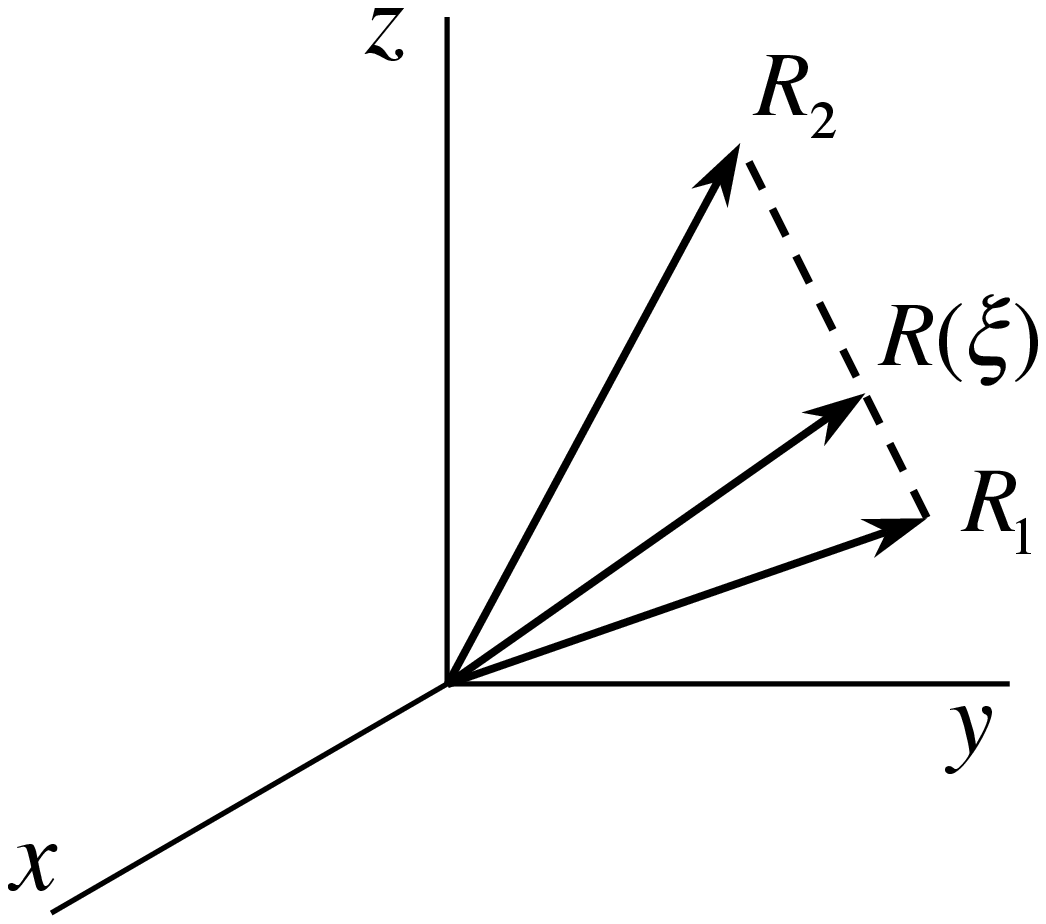}\includegraphics[width=2.7in]{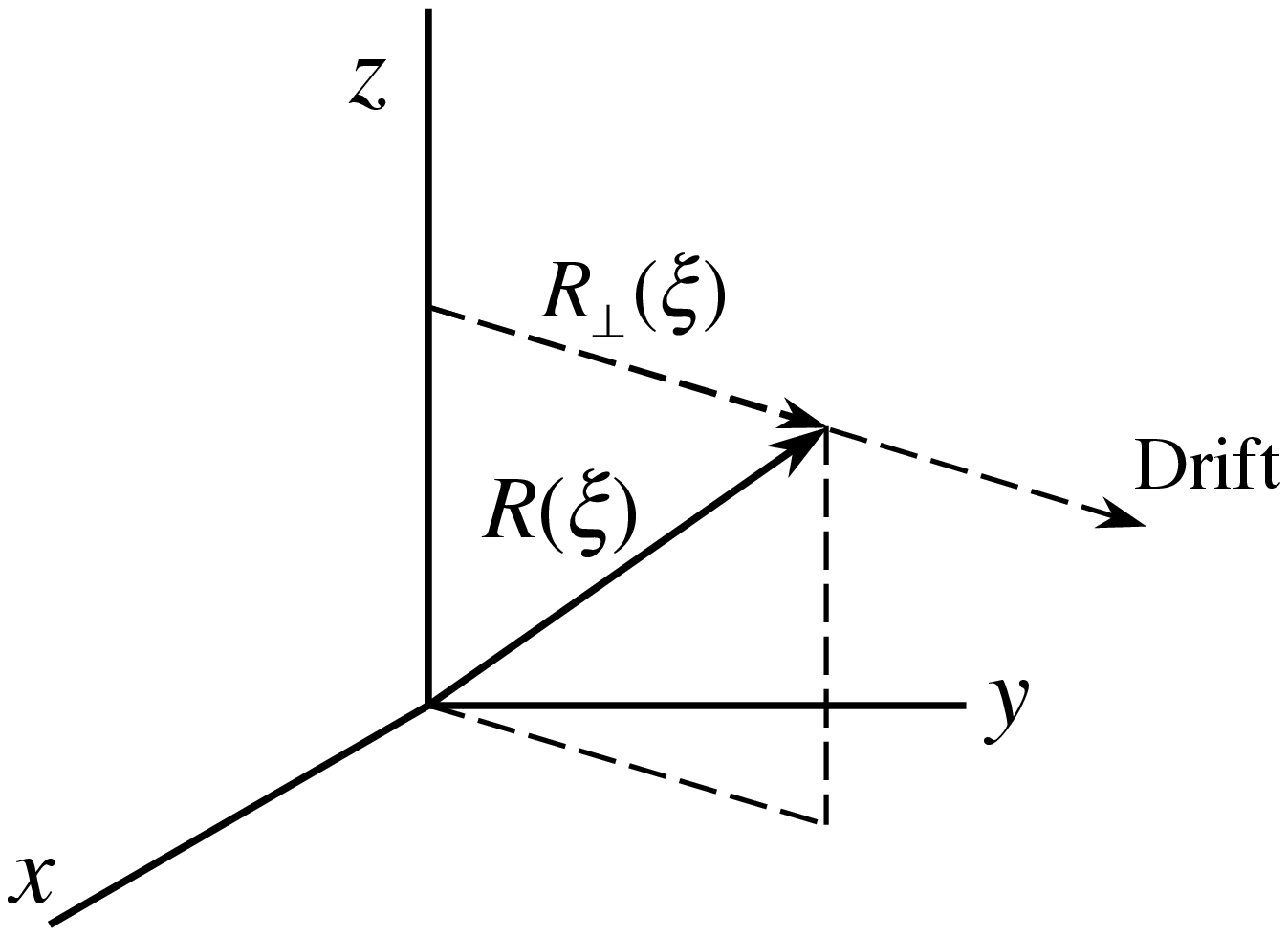}
\caption{\label{fig:line-drift} Left (a): Track segment $({\bf R}_1,{\bf R}_2)$ with typical starting point ${\bf R}(\xi)$.  Right (b): Details of a typical starting point showing its decomposition ${\bf R}(\xi)={\bf R}_\perp(\xi)+R_\parallel {{\bf\hat z}}$ into a piece along the drift direction and a piece parallel to the axis of the cylinder.}
\end{center}
\end{figure}

To illustrate the possible use of our results, instead of a single
point source arising at ${\bf r} = {\bf R}_{0} $ at $t = 0$, we
consider the ionization along the path of a relativistic charged
particle through a cylindrical drift chamber.  The ionization trail is
assumed linear (in both senses) and is, on the time scale relevant for
the subsequent diffusion, deposited instantaneously. The resulting
charge distribution is thus a superposition of distributions of the
form Eq.(\ref{eq:ansatztwo}) for contributions along the particle's
track. For definiteness we consider a track segment with end points
${\bf R}_{1}, \ {\bf R}_{2} $, as shown in figure
(\ref{fig:line-drift}a), with a typical point on the track labeled by
${\bf R}(\xi)$.  For convenience, define the distance between the end
points as $L = |{\bf R}_{2} - {\bf R}_{1}| $.  Then we have ${\bf
  R}(\xi) = {\bf R}_{1} + (\xi/L) ({\bf R}_{2} - {\bf R}_{1}) $.  In
figure (\ref{fig:line-drift}b), the typical starting point ${\bf
  R}(\xi)$ is singled out.  We define $ R_{\parallel}(\xi) = \hat{{\bf
    z}} \cdot {\bf R}(\xi) $ and ${\bf R}_{\perp}(\xi) = {\bf R}(\xi)
- \hat{{\bf z}} R_{\parallel}(\xi) $.  Then the drift center's
coordinate ${\bf r}_{0}(t) $ can be written according to
Eq.(\ref{eq:r_of_t}) as
\begeq
{\bf r}_{0}(t) \ = \ {\bf R}_{\perp}(\xi) [1 + t/T(\xi) ]^{1/2} \ + \ \hat{{\bf z}} R_{\parallel}(\xi) 
\end{equation}
where from (30), since $x_{0}(\xi) = |{\bf R}_{\perp}(\xi)| $, we have
\begeq
 T(\xi) = \frac{|{\bf R}_{\perp}(\xi)|^{2}}{2 \mu A}
\end{equation}
In the exponent of $\psi$ the Cartesian components of the coordinate difference, $({\bf r} - {\bf r}_{0}(t))$, must be taken with respect to ${\bf R}_{\perp}(\xi) $ as the $x$-direction:
\begin{eqnarray}
\rho_1\equiv({\bf r} - {\bf r}_{0}(t))_{1} &  = & ({\bf r} - {\bf r}_{0}(t)) \cdot \hat{{\bf R}}_{\perp}(\xi) \\
\rho_2\equiv({\bf r} - {\bf r}_{0}(t))_{2} & =  & ({\bf r} - {\bf r}_{0}(t)) \cdot (\hat{{\bf z}} \times \hat{{\bf R}}_{\perp}(\xi))\\
\rho_3\equiv({\bf r} - {\bf r}_{0}(t))_{3} & =  &z-R_\parallel(\xi)
\end{eqnarray} 

\indent The normalized diffused charge density in space and time for the track segment can thus be written as a numerical integral over $\psi$ given by Eq.(\ref{eq:ansatztwo}),
\begeq
\Psi({\bf r},  t| {\bf R}_{1}, {\bf R}_{2}) \ = \ \frac{1}{L}\int_{0}^{L} d\xi \ \psi({\bf r}, {\bf R}(\xi), t)
\end{equation}
The functions $g_{1}(s)$ and $g_{2}(s)$ are to be evaluated at $s = t/T(\xi)$; $g_3(s)=0.$\\

Acknowledgments: This work was supported by the Director, Office
of Science, High Energy Physics, U.S. Department of Energy under
Contract No. DE-AC02-05CH11231. We thank David Nygren for drawing our
attention to this class of ionic diffusion problems.

\appendix
\section{Solution of Ricatti equations for cylindrical geometry}
  The procedure for solution of the Riccati equations can be found in Refs.{\cite{korn,riccati}.
 The general form is
\begeq
y'= q_0(x) +q_1(x)y +q_2(x)y^2.
\endeq
To find the solution we consider the associated second order homogeneous linear differential equation
\begeq
u'' -P(x) u' +Q(x)u=0
\endeq
with
\begeq
Q=q_2 q_0;\quad P=q_1 +\frac {q_2'}{q_2}.
\endeq
Then the solution to the original equation is
\begeq
y =-\frac {u'}{q_2u}.\label{eq:original}
\endeq

(a) For $g_1$ we have
\begeq
q_2=-\frac 1s;\qquad  q_1=-\frac1{s(1+s)};\qquad q_0=\frac 1{1+s};
\endeq
so
\begeq
Q=-\frac 1{s(1+s)};\qquad P=-\frac{2+s}{2(1+s)};
\endeq
and
\begeqar
u''+\frac{2+s}{s(1+s)}u'-\frac1{s(1+s)}u&=&0,\non
s(1+s) u'' +(2+s) u' -u&=&0.
\endeqar
We recognize this as the hypergeometric equation for a function $w(z)$\cite{ref:NBS},
\begeq
z(1-z)\frac{d^2w}{dz^2}+[c-(a+b+1)z]\frac{dw}{dz}-abw=0,\label{eq:hyper}
\endeq
whose solution is
\begeq
F(a,b;c;z)= 1+\frac {ab}c z+\frac {a(a+1)b(b+1)}{c(c+1)}\frac{z^2}{2!}+\ldots.
\endeq
In particular, we have
\begeq
u_1=F(1,-1;2;-s)=1+\frac 12 s
\endeq
and finally, from Eq.(\ref{eq:original}), we have
\begeq
g_1=\frac {s}{2+s}.
\endeq
(b) For $g_2$ we have
\begeq
q_2=-\frac 1s;\qquad  q_1=-\frac{2s+1}{s(1+s)};\qquad q_0=-\frac 1{1+s},
\endeq
so
\begeq
Q=\frac 1{s(1+s)};\qquad P=-\frac{2+3s}{2(1+s)};
\endeq
and
\begeqar
u''+\frac{2+3s}{s(1+s)}u'+\frac1{s(1+s)}u&=&0,\non
s(1+s) u'' +(2+3s) u' +u&=&0.
\endeqar
with the solution
\begeq
u_2=F(1,1;2;-s)=-\frac 1s\ln(1+s),
\endeq
from which we find, from Eq.(\ref{eq:original})
\begeq
g_2=-1 +\frac s{(1+s)\ln(1+s)}.
\endeq
This can be confirmed by direct substitution in the original differential equation.
\section{Example of spherical geometry}
The volume consists of the region between two concentric conducting spherical shells of radii $a, \ b \ (a < b)$. Voltages are applied to the shells  such that a radial electric field ${\bf E} = A {\bf r}/r^{3} $ is produced.  It is assumed that the force on the ions is outward, unless stated to the contrary.\\
Consider the starting point to be at ${\bf R} = (R,\ 0,\ 0)$ at $t = 0$, with $a < R < b$. The drifting center moves outward along the $x = x_{1}$ axis.  The classical equation of motion, ${\bf v} = \mu \ {\bf E}$, reduces to $dx/dt \ = \ \mu A/x^{2}$, with solution, 
\begeq
r_{0}(t) \ = \ (R^{3} \ + \ 3 \mu At)^{1/3}
\end{equation}
We define
\begeq
T \ = \ \frac{R^{3}}{3 \mu A}\label{eq:t_spherical}
\end{equation}
Then we have
\begeq
r_{0} \ =\ R(1 \ +\ \frac{t}{T})^{1/3}
\end{equation}

The needed partial derivatives of ${2\mu \ \bf E}$  near the $x$ -axis are
\[2\mu \ \partial_{1}E_{1} = \ -\ \frac{2}{3T}\cdot\frac{1}{(1 \ +\ t/T)}\]
\[ 2\mu \ \partial_{2}E_{2} = \ +\ \frac{1}{3T}\cdot\frac{1}{(1 \ +\ t/T)}\]
with the derivative in the $x_{3}$ direction the same as for $x_{2}$.  The equations for $g_{j}, \ j=1, \ 2$ take the form
\begeq
\frac{dg_{j}}{dt} \ + \ \frac{g_{j}(1+g_{j})}{s} \ + \alpha \ \frac{(1+g_{j})}{(1+s)} \ = \ 0
\end{equation}
where, as before, $s = t/T $.  The parameter $\alpha$ is $\alpha = -2/3, \ +1/3$ for the longitudinal $(j=1)$ and transverse $(j=2)$ directions, respectively.\\
With the introduction of $g_{j} =  s \cdot w(s)$, the equation for $g_{j}$ becomes a Ricatti  equation for $w(s)$:
\begeq
\frac{dw}{ds} \ + \ w^{2} \ + \ w(\frac{2}{s} + \frac{\alpha}{1+s}) \ +\ \frac{\alpha}{s(1+s)} \ =\ 0.
\end{equation}
With $z = -s$ and $w= \frac{1}{y}\frac{dy}{ds}$, this becomes the hypergeometric equation, Eq.(\ref{eq:hyper}), for $F(a,b;c;z)$ with $c=2, \ a+b \ = 1 + \alpha, \ ab = \alpha$. Thus for $j = 1$, $ a= 1,\ b = -2/3, \ c = 2$ and for $j = 2$, $a = 1, \ b = 1/3, \ c = 2 $.  With some algebra and use of relations of the hypergeometric functions, we find the solutions for the longitudinal and transverse functions,  $g_{1}$ and $g_{2}$ :
\begeq
g_{1} \  =  \  \frac{1}{3}\cdot \frac{s}{(1+s)} \cdot \frac{F(1, \ 1/3;\  3 ; \ s/(1+s) \ )}{F(1; \ -2/3; \ 2; \ s/(1+s) \ )}
\end{equation}
\begeq
g_{2} \   =  \  \frac{-1}{6} \cdot \frac{s}{(1+s)}\cdot\frac{F(1, \ 4/3; \ 3 ; \ s/(1+s) \ )}{F(1; \ 1/3;\ 2;\  s/(1+s) \ )}
\end{equation}
The classical center is given by $r_{0}(t)  =  R(1+s)^{1/3}$. For example, for 
$r_{0}(t)/R = 2$, we have $s = 7$.\\

Figures \ref{fig:sphere_g1g2} and \ref{fig:sphere_sigma} show the
behavior of $g_{1}(s)$ and $ - \ g_{2}(s)$ versus $s$ and the
corresponding widths, $\sigma_{\parallel} / \sigma_{0}$ and
$\sigma_{\perp} / \sigma_{0}$, qualitatively similar to the
cylindrical situation, but with differences stemming from the more
rapid fall-off of the electric field with distance.

\begin{figure}[t]\begin{center}
\includegraphics[width=4in,angle=90]{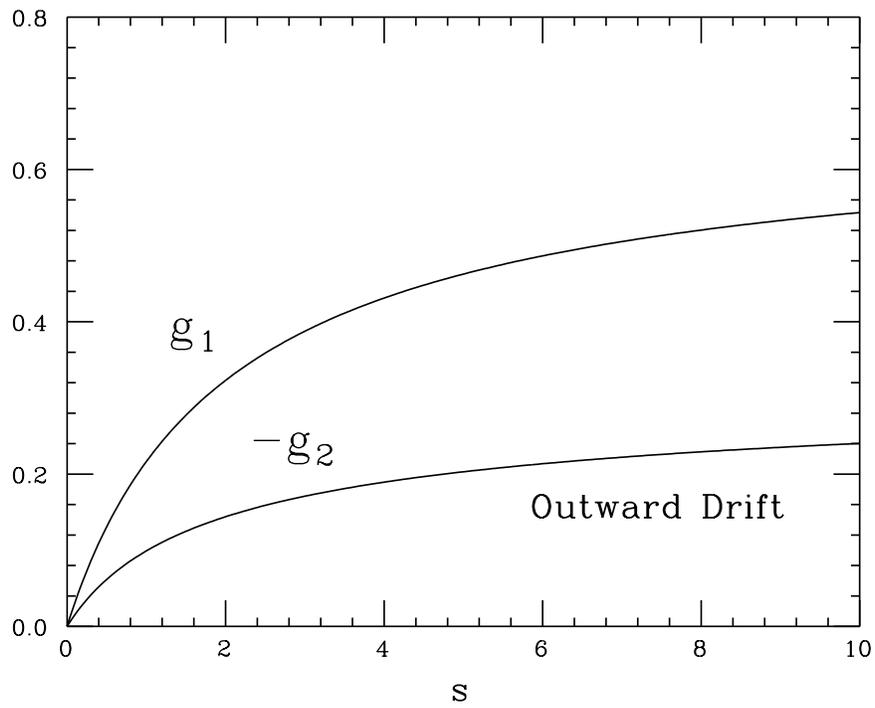}
\caption[99]{The functions appearing in the Gaussian description of $\psi$ for spherical geometry and outward drift
as functions of the dimensionless time variable $s=t/T$, where $T$ is now given by Eq.(\ref{eq:t_spherical}).} \label{fig:sphere_g1g2}
\end{center}
\end{figure}
\clearpage
\begin{figure}[t]\begin{center}
\includegraphics[width=4in,angle=90]{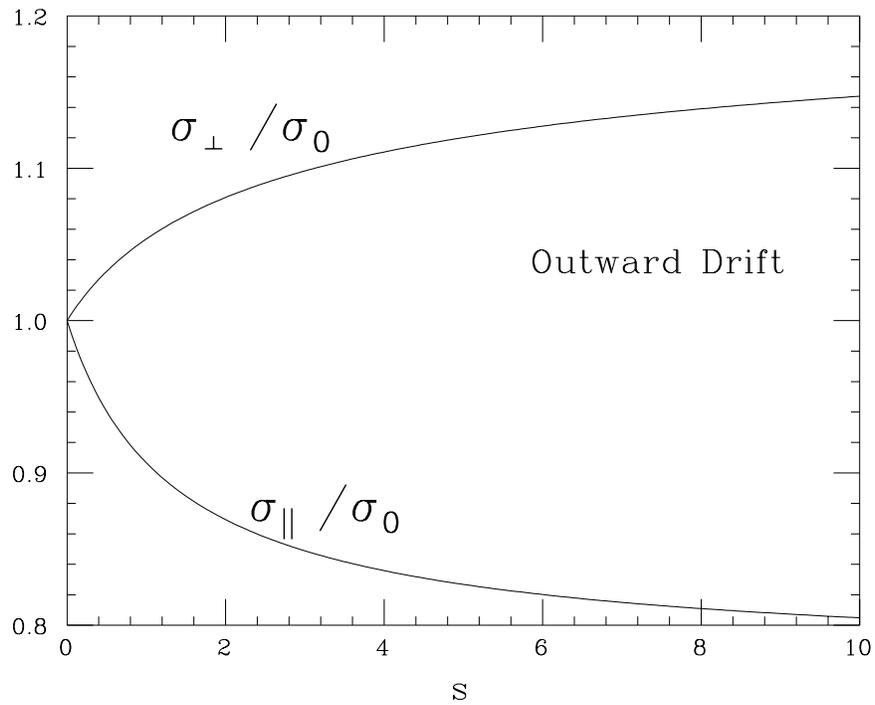}
\caption[99]{The effective dimensionless widths in the directions
  parallel and perpendicular to the radial direction for spherical
  geometry and outward drift as functions of the dimensionless time
  variable $s=t/T$, where $T$ is now given by
  Eq.(\ref{eq:t_spherical}).} \label{fig:sphere_sigma}\end{center}
\end{figure} 

\clearpage


\clearpage
\begin{thebibliography}{99}
\bibitem {Huxley} L. G. H. Huxley and R. W. Crompton, \emph{The Diffusion and Drift of Electrons in Gases}, Wiley, New York, 1974.
\bibitem {Mason} E. A. Mason and E. W. McDaniel, \emph{Transport Properties of Ions in Gases}, Wiley, New York, 1988.
\bibitem{Ellis_1} H. W. Ellis, R. Y. Pai, E. W. McDaniel, E. A. Mason, and L. A. Viehland, \emph{Transport properties of gaseous ions over a wide energy range}, At.
 Data Nucl. Data Tables {\bf 17} (1976) 177-210.
\bibitem{Ellis_2} H. W. Ellis, E. W. McDaniel, D. L. Albritton, L. A. Viehland, S.
 L. Lin, and E. A. Mason, \emph{Transport properties of gaseous ions over a wide energy range: Part 2}, At. Data Nucl. Data Tables {\bf 22} (1978) 179-217.
\bibitem{Ellis_3} H. W. Ellis, M. G. Thackston, E. W. McDaniel, and E. A. Mason, \emph{Transport properties of gaseous ions over a wide energy range: Part 3}, At. Data Nucl. Data Tables {\bf 31} (1984) 113-151.
\bibitem{notation} Td stands for Townsend. $1 \  \mathrm{Td} = 10^{ -17} {\rm V-cm
}^{2}$.  $E$ is the electric field in volts/cm; $N$ is the number of fluid atoms per cubic centimeter.
\bibitem{ref:NBS} M. Abramowitz and I.A. Stegun, eds., \emph{Handbook of Mathematical Functions},
National Bureau of Standards, Washington, 1964, p. 562, Eq. 15.5.1.
\bibitem{korn}G. A. Korn and T. M. Korn, {\it Mathematical Handbook for Scientists
 and Engineers}, 2nd. ed, McGraw-Hill, N. Y., 1968, pp. 250-251.
\bibitem{riccati}\url{http://en.wikipedia.org/wiki/Riccati_equation}.\end{thebibliography}
\end{document}